\documentclass{rspublic}
\usepackage{amsmath}
\usepackage{amsfonts}
\usepackage{amssymb}
\usepackage[Symbol]{upgreek}
\usepackage[nointegrals]{wasysym}
\usepackage{mathrsfs}
\usepackage{graphicx}
\begin{document}
\title[Navier-Stokes Equations in Simple Geometry]{\large On Exact Solutions of the Navier-Stokes Equations for Uni-directional Flows}
\author[F. Lam]{F. Lam}
\label{firstpage}
\maketitle
\begin{abstract}{Navier-Stokes Equations; Viscosity; Pipe Hagen-Poiseuille Flow; Plane Couette Flow; Plane Poisueille Flow; Reynolds Number;}

In the present note, we show that the uni-directional flows in a rectangular channel and in a circular pipe are exact spatio-temporal solutions of the Navier-Stokes equations over a short time interval. We assert that the classical plane Poiseuille-Couette flow and Hagen-Poiseuille flow are time-independent approximations of the exact solutions if an appropriate initial velocity distribution at starting location is specified. Conceptually, there do not exist absolute steady flows starting from unspecified initial data. The classic experimental measurements by Poiseuille can be explained in terms of the evolutional solutions. In particular, the pipe flow does not have a time-independent characteristic velocity. The orthodox notion that the parabolic profile exists for arbitrary Reynolds numbers is unwarranted.
\end{abstract}
%
%
\section{Introduction}
In the Eulerian description of the motion of an incompressible, homogeneous Newtonian fluid, the momentum and the continuity equations for fluid dynamics are 
\begin{equation} \label{ns}
	{\partial {\bf u}}/{\partial t} + ({\bf u} . \nabla) {\bf u} = \nu \Delta {\bf u}   - {\rho}^{-1} \nabla p, \;\;\; \nabla . {\bf u} =0,
\end{equation}
where the velocity vector ${\bf u}={\bf u}({\bf x},t)$ has the components $(u,v,w)$, the scalar quantity $p=p({\bf x},t)$ is the pressure, the space variable is denoted by ${\bf x}=(x,y,z)$, and $\Delta$ is the Laplacian. The density and the viscosity of the fluid are denoted by $\rho$ and $\mu$ respectively. The kinematic viscosity is $\nu =\mu/\rho$.  These equations are known as the Navier-Stokes equations. They are derived on the basis of the continuum hypothesis. 

We seek the solution of (\ref{ns}) as an initial-boundary value problem in the space-time domain denoted by $\Upomega {\times} [0,t>0]$, where $\Upomega$ has a smooth impermeable boundary $\partial \Upomega$. 
The initial condition is given by
\begin{equation} \label{ic}
 {\bf u}({\bf x},t{=}0) = {\bf u}_0({\bf x})\;\;\; {\bf x} \in \Upomega,
\end{equation} 
and the no-slip boundary condition is
\begin{equation} \label{bc}
 {\bf u}({\bf x},t) = 0\;\;\; {\bf x} \in \partial \Upomega.
\end{equation} 

The equations of motion (\ref{ns}) are often re-written in dimensionless form
\begin{equation} \label{ns-d}
	{\partial {\bf u^*}}/{\partial t^*} + ({\bf u^*} . \nabla) {\bf u^*} = \frac{1}{Re} \Delta {\bf u^*}   - \nabla p^*, \;\;\; \nabla . {\bf u^*} =0,
\end{equation}
where $Re$ denotes the Reynolds number with reference to a characteristic length scale and a characteristic velocity. A prevailing concept is that there exist some basic flows, such as plane Poiseuille flow and the Hagen-Poiseuille flow in a circular pipe, which are ``exact" solutions of the Navier-Stokes equations, though these basic flows are in fact the solutions of {\it simplified} equations of motion. Nevertheless, one important implication is that the basic flows exist at all Reynolds numbers. In fact, the stability of a flow depends essentially on this idea which quantifies the growth or decay of disturbances as the Reynolds number varies, as a parameter. In the present note, we show that approximating or simplifying the equations of motion without due justifications can be very misleading, as the solutions of the approximated equations are inconsistent in the spatio-temporal evolution of fluid motion.
\section{Some particular solutions}
Let us consider plane Poiseuille flow in a rectangular channel with half width $b$. 
As in many textbooks on fluid mechanics, we make the following assumptions for the flow:
\begin{equation} \label{ppf-cond}
v=w=0,\;\;\; \partial u/\partial t = 0,\;\;\; u =u(y),\;\;\; \mbox{and} \;\;\; p =p(x).
\end{equation}
The momentum equation becomes an ordinary differential equation 
\begin{equation} \label{ppfns}
	{\rd^2 u}/{\rd y^2} =  (\partial p/ \partial x)/\mu, \;\;\; u(\pm b)=0.
\end{equation}
The {\it time-independent steady} solution of the boundary value problem is given by 
\begin{equation} \label{ppf}
u(y)=- \frac{1}{2 \mu} \Big( \frac{\partial p}{\partial x} \Big)\: \big( b^2 - y^2 \big).
\end{equation}

Alternatively, we seek Navier-Stokes solutions by assuming
\begin{equation} \label{ghost-cond}
v=w=0, \;\;\; u=u(t),\;\;\; \mbox{and} \;\;\; \partial p/\partial x =\partial p/\partial x(t).
\end{equation}
Now the flow evolution is an initial-value problem:
\begin{equation} \label{ghostns}
	{\rd u}/{\rd t} = - (\partial p/ \partial x)/\rho, \;\;\; u(t=0)=u_0
\end{equation}
Its solution defines a class of {\it space-independent unsteady} flows,
\begin{equation} \label{ghost}
u(t)=u_0 - \frac{1}{\rho}\int_0^t \Big( \frac{\partial p}{\partial x} \Big)(\tau) \rd \tau,
\end{equation}
which complements solution (\ref{ppf}) by space-time symmetry.

Both solutions (\ref{ppf}) and (\ref{ghost}) may be regarded as exact particular solutions of the Navier-Stokes equations, only in a restrictive sense. Moreover, plane Poiseuille flow is considered to exist at arbitrary Reynolds number $Re_b=Ub/\nu$, where $U$ is the velocity at the channel centre and $\nu>0$. Similarly, we claim that the unsteady solution (\ref{ghost}) is valid for all unit Reynolds number $Re=1/\nu$.

Strictly speaking, the correctness of the profile (\ref{ppf}) must be tested experimentally by intelligent beings squatting absolutely on space while supernaturals crawling in time may be able to verify the flow (\ref{ghost})!

For a straight circular pipe of radius $R$, we have the analogous expression to (\ref{ppf}): 
\begin{equation} \label{hpf}
u(r)=- \frac{1}{4 \mu} \Big( \frac{\partial p}{\partial x} \Big)\: \big( R^2 - r^2 \big).
\end{equation}
This is known as the Hagen-Poiseuille flow.

In fact, we have no {\it a priori} justifications to detach space from time or vice versa, as implied in the simplification (\ref{ppf-cond}) or (\ref{ghost-cond}). Every fluid motion develops in space {\it as well as} in time. Since we have separately considered the initial-boundary data (\ref{ic}) and (\ref{bc}), neither (\ref{ppf}) nor (\ref{ghost}) can be regarded as correct descriptions of the evolution. Practically, the velocity profile (\ref{hpf}) is accepted as ``correct" because there are experimental verifications (Hagen 1839; Poiseuille 1840) even though these experiments ought to have been carried out {\it locally}. Within the experimental errors, the parabolic profile agreed with the measurements of ensemble averages. The experiment repeats and the data sampling must have been undertaken over time, though the time-dependence may not be shown in an explicit formulation.

We recall that the first theoretical explanation of (\ref{hpf}) in circular tubes was not done by solving equations but was derived by Poiseuille by considering the momentum balance (see, for example, \S 331 of Lamb 1975) over a fully developed  flow or a transient with no reference to its initial condition. The question is, to what extent, can we ignore the temporal effects?
\section{Flow evolution over space-time}
Instead of (\ref{ppf-cond}), we postulate 
\begin{equation} \label{ts}
\forall t \in [t_s,t_s+\epsilon], \; \epsilon>0, \;\;\; v=w=0, \;\;\; u=u(y,t),\;\;\; \mbox{and} \;\;\; p=p(x,t),
\end{equation}
where $t_s$ stands for starting time, and $\epsilon$ denotes a short interval. The start refers to either the initialisation of a motion from rest or a subsequent moment when the assumed uni-directional flow has been established at least over a part of the apparatus. The second option is more relevant to experimental measurements because the flow at the inlet may be specified as initial data. With these understandings, we set $t_s=0$.
\subsection{Plane Poiseuille flow}
We consider the following initial-boundary value problem:
\begin{equation} \label{ppfns-ibp}
\partial u/\partial t - \nu {\partial^2 u}/{\partial y^2} = q(x,t), \;\;\; u(y,0)=u_0(y),\;\;\; u(0)=u(1)=0.
\end{equation}
where $q(x,t){=-}(\partial p/ \partial x)/\rho$ is the scaled pressure gradient ($b=1$). The solution of this diffusion problem is well-known, (see, for example, Carslaw \& Jaeger 1947),
\begin{equation} \label{ppf-ibp}
u(y,t)=\int_0^1 u_0(\sigma) H(y,\sigma,t) \rd \sigma + \int_0^t \int_0^1 q(x,\tau) H(y,\sigma,t{-}\tau) \rd \sigma \rd \tau, 
\end{equation}
where Green's function is given by
\begin{equation} \label{ppf-green}
H(y,\sigma,t)=2\sum_{n=1}^{\infty} \; \exp \big(- n^2 \pi^2 \nu t \big) \sin\big(n \pi y\big)\sin\big(n \pi \sigma \big),
\end{equation}
for all time $t \in [0,\epsilon]$, and in general $\epsilon=\epsilon(u_0,q)$. For constant pressure gradient $q(x)=q_0$, the second integral in (\ref{ppf-ibp}) is found to be
\begin{equation} \label{ppf-q0}
\frac{4 q_0}{\nu} \sum_{\mbox{{\scriptsize odd}}\;n}^{\infty} \frac{\sin(n \pi y)}{n^3 \pi ^3}\Big( 1 - \exp\big(-n^2 \pi^2 \nu t \big) \Big).
\end{equation}

Consider the particular initial profile at a location closed to the inlet
\begin{equation} \label{id}
u_0(y)=\alpha y(1-y),
\end{equation}
where $\alpha$ is a given constant, its contribution to the solution is given by
\begin{equation} \label{ppf-id}
8 \alpha \sum_{\mbox{{\scriptsize odd}}\;n}^{\infty}\frac{\sin(n \pi y)}{n^3 \pi^3} \: \exp\big(- n^2 \pi^2 \nu t \big).
\end{equation}

We make the following assertions:
\begin{itemize}
\item In view of (\ref{ppf-q0}) and (\ref{ppf-id}), the subsequent flow development of initial data (\ref{id}) is steady over a {\it short time interval} if
\begin{equation*}
\alpha = q_0/(2\nu).
\end{equation*}
The solution can be expressed as
\begin{equation*}
u(y,t) \sim u(y) =  - \frac{1}{2 \mu} \Big(\frac{\partial p} {\partial x}\Big)\:\sum_{\mbox{{\scriptsize odd}}\;n}^{\infty} \frac{\sin(n \pi y)}{n^3 \pi ^3}=- \frac{1}{2 \mu} \Big(\frac{\partial p} {\partial x}\Big)\;(y-y^2),
\end{equation*}
(cf. (\ref{ppf})), by virtue of the Fourier series expansion for functions $y$ and $y^2$ over $0 \le y < 1$.

\item For $\alpha\neq -({\partial p}/{\partial x})/(2 \mu)$ in (\ref{id}), the flow is time-dependent. However, the following approximation is valid. Within a short time $t \rightarrow 0^+$ or for small viscosity $\nu$ such that the exponential decay has virtually no influence. Measurements made at a small distance downstream must be very close to the initial data, as implied in (\ref{ppf-id}). The small viscosity requirement is relevant to many experiments using air ($\nu \sim O(10^{-5})$ in SI units, $m^2s^{-1}$) and pure water ($\nu \sim O(10^{-6})$) under the conditions of standard temperature and pressure. 

\item For an arbitrary initial velocity $u_0$ of a polynomial satisfying the boundary conditions, the flow is unsteady in general. If we ignore the contribution from the pressure gradient over the short-time interval $\epsilon$, the flow downstream of the inlet resembles the initial data, in the limit of $\nu t \rightarrow 0$, according to Fourier series expansions for simple functions.
\end{itemize}
\subsection{Plane Couette flow}
If we consider the following modification of the boundary conditions in (\ref{ppfns-ibp}):
\begin{equation} \label{pcfns-ibp}
u(y=0)=A(t),\;\;\;u(y=1)=B(t),
\end{equation}
we obtain the solution for plane Couette flow:
\begin{equation} \label{pcf-ibp}
\begin{split}
u(y,t)=\int_0^1 & u_0(\sigma) H(y,\sigma,t) \rd \sigma + \int_0^t \int_0^1 q(x,\tau) H(y,\sigma,t{-}\tau) \rd \sigma \rd \tau \\
& + \int_0^t (A-B)(\tau)\: h(y,t{-}\tau) \rd \tau,
\end{split}
\end{equation}
where
\begin{equation} \label{pcf-green}
h(y,t)=2\sum_{n=1}^{\infty} \; (-1)^{n} n \pi \: \exp \big(- n^2 \pi^2 \nu t \big) \sin\big(n \pi y\big).
\end{equation}
If the boundary conditions are time-independent, the last integral can be evaluated as
\begin{equation} \label{pcf-bdy}
\frac{2(A-B)}{\nu} \sum_{n=1}^{\infty} (-1)^n \frac{\sin(n \pi y)}{n \pi}\Big( 1 - \exp\big(-n^2 \pi^2 \nu t \big) \Big).
\end{equation}
Let us consider several particular cases of our general solution:

\begin{itemize}
\item An impulsively started motion $u_0{=}0$, with zero pressure gradient and boundary conditions $A=0$, $B/\nu=+1$. 

In view of (\ref{pcf-ibp}) and (\ref{pcf-bdy}), the time-independent solution of this motion is plane Couette flow,
\begin{equation} \label{pcf-imp}
u(y,t)\sim u(y) = 2 \sum_{n=1}^{\infty} (-1)^{n+1} \frac{\sin(n \pi y)}{n \pi} = y,\;\;\;0 \le y < 1.
\end{equation}
Evidently, the viscosity plays a key role in determining how soon the temporal effects may be neglected.

\item A motion driven by the upper boundary $B = Q \sin(\omega t)/2$ ($A{=}0$) with initial data $u_0{=}0$ and zero pressure gradient.

The solution is given by
\begin{equation} \label{pcf-ivp}
u(y,t) = Q  \sum_{n=1}^{\infty}(-1)^{n+1} \:\frac{(n \pi) \: \sin(n \pi y)}{\omega^2+ n^4 \pi^4 \nu^2} \: \chi(\omega),
\end{equation}
where
\begin{equation*}
\chi(\omega)=(n^2 \pi^2 \nu)\: \sin(\omega t) - \omega \cos(\omega t) + \omega \exp\big(- n^2 \pi^2 \nu t \big).
\end{equation*}
In practice, this result may only be realised within a short period of time after the sinusoidal force has been applied since the non-linearity cannot be ignored, especially for large $Q$.

\item A motion driven by $A = \alpha \sin(\omega_{\alpha} t)/2$ and $B = \beta \sin(\omega_{\beta} t)/2$ with zero initial data and zero pressure gradient.

In view of the last solution, we find
\begin{equation} \label{pcf-control}
u(y,t) = \sum_{n=1}^{\infty}(-1)^{n+1} \:{(n \pi) \: \sin(n \pi y)} \Big(\: \frac{\alpha \chi(\omega_{\alpha})}{\omega_{\alpha}^2 + n^4 \pi^4 \nu^2}- \frac{\beta \chi(\omega_{\beta})}{\omega_{\beta}^2 + n^4 \pi^4 \nu^2} \:\Big).
\end{equation}
The solution can be considered as a model for simulating the effect of wall vibrations in experiments if we choose small amplitudes $\alpha$ and $\beta$.
\end{itemize}

More complicated flows can be generated and analysed if we include non-zero initial data and time-dependent pressure gradients.
\section{Hagen-Poiseuille flow}
In a cylindrical co-ordinates system $(r, \theta, z)$, consider $({\bf u},p)=({\bf u},p)(r, \theta, z, t)$. The velocity components in the radial and circumferential directions are hypothesised as identically zero over time $t \in [0,\epsilon]$. The axial component is treated as a function of the independent variables $r$ and $t$, 
\begin{equation*}
w=w(r,t),\;\;\;p=p(z,t).
\end{equation*}
By analogy, the differential equation for Hagen-Poiseuille flow in a circular pipe of radius unity becomes (see, for example, Batchelor 1973)
\begin{equation} \label{hpfns-ibp}
\begin{split}
\frac{\partial w} {\partial t} - \nu &\Big( \frac{\partial^2 w}{\partial r^2} + \frac{1}{r} \frac{\partial w}{\partial r} \Big) = \nu \psi(z,t),\\
&w(r,0) =w_0(r),\;\;\; w(1)=0,
\end{split}
\end{equation}
where we seek bounded solution, $|w(r)| {<} \infty, \:\forall r {\in} [0,1)$, and $\psi=-(\partial p/ \partial z)/\mu$ denotes the axial pressure gradient. The solution of the initial-boundary value problem can be expressed in terms of the Bessel functions of orders zero and one,
\begin{equation} \label{hpf-ibp}
w(r,t)=\int_0^1 w_0(s) G(r,s,t) \rd s + \nu \int_0^t \int_0^1 \psi(z,\tau) G(r,s,t{-}\tau) \rd s \rd \tau,
\end{equation}
where Green's function is given by
\begin{equation} \label{hpf-green}
G(r,s,t)=2 \sum_{n=1}^{\infty} \; \exp \big(- \lambda^2_n  \nu t \big) \: s \frac{J_0 \big(\lambda_n r \big) J_0 \big(\lambda_n s \big)}{J^2_1 \big(\lambda_n  \big)},
\end{equation}
where the constants, $\lambda_n, n=1,2,\cdots,$ are the positive zeros of $J_0(\lambda_n)=0$. The first few of them are
\begin{align*}
\lambda_1&=2.40482556,&\lambda_2&=5.52007811,&\lambda_3&=8.65372791,&\lambda_4&=11.79153444,\\
\lambda_5&=14.93091771,&\lambda_6&=18.07106397,&\lambda_7&=21.21163663,&\lambda_8&=24.35247153,\\
\lambda_9&=27.49347913,&\lambda_{10}&=30.63460647,&\lambda_{11}&=33.77582021,&\lambda_{12}&=36.91709835.
\end{align*}
The other zeros may be calculated by making use of the asymptotic formula
\begin{equation*}
	\lambda_n = \beta + 1/(8 \beta),\;\;\; \beta=(n-1/4)\pi.
\end{equation*}
In particular, none of these zeros coincides with any zero of $J_1(\sigma_n)$. Bessel functions $J_n(x)$ are entire functions of the argument $x$ (Watson 1944),
\begin{equation*}
J_n(x) = \Big(\frac{x}{2}\Big)^n \sum_{k=0}^{\infty} \frac{(-x^2/4)^k}{k! \:\Gamma(k+n+1)},\;\;\;n\geq0.
\end{equation*}
Explicitly,
\begin{equation*}
	J_0(x) = 1 - \frac{x^2}{2^2}+\frac{x^4}{2^2\:4^2}-\frac{x^6}{2^2\:4^2\:6^2}+ \cdots.
\end{equation*}
We notice that
\begin{equation*}
	\max|J_0| = 1 \;\;\; \mbox{at} \;\;\;x=0, \;\;\;\max|J_1| \approx 0.5819 \;\;\; \mbox{at} \;\;\;x \approx 1.8412.
\end{equation*}
As $x$ tends to $0$, $J_1(x)$ tends to $0$, and $J_1(\lambda_1) \approx 0.5191$.
 
Evidently, the flow evolution reaches a steady state in the interval $\epsilon$ if $\nu t \rightarrow 0^+$ when $\exp(-\lambda^2_n \nu t) \sim O(1)$ for some $n \leq 10 $ (say). The smallness assumption for $\nu t$ may be justified for air and water. Practically, the uni-directional flow can only be observed during an interval $t \sim O(1)$ measured in seconds under normal laboratory conditions. Thus the temporal effects due to the exponential decay reduces to the extent when any experimental measurements of the flow become time-insensitive. 

In view of the identity,
\begin{equation*}
\int_0^1 s (1-s^2)^{m-1} J_0(\lambda_n s) \rd s = \frac{2^{m-1}(m-1)!}{\lambda^{m}_n} \; J_m(\lambda_n),
\end{equation*}
the integral involving the Bessel functions in Green's function may be evaluated, particularly for the initial condition having a ``parabolic" distribution.

Assuming a time-independent pressure gradient, the second term in (\ref{hpf-ibp}) contributes nothing as $\nu t \rightarrow 0$. Thus the solution due to the initial data,
\begin{equation*}
w_0(r) = \beta(1- r^2), \;\;(\beta = \const)
\end{equation*}
is given by  
\begin{equation} \label{hpf-t}
\begin{split}
w(r,t)\sim w(r) & = 4 \beta \sum_{n=1}^{\infty} \; \frac{J_2(\lambda_n)}{\lambda^2_n\;J_1^2(\lambda_n)}  \; J_0(\lambda_n r) \\
& \approx \beta \sum_{n=1}^{m} \; \frac{J_2(\lambda_n)}{J_1^2(\lambda_n)} \; \Big( \frac{4}{\lambda^2_n} - r^2 + \frac{\lambda^4_n}{16} r^4 - \cdots \Big)\\
& \approx \beta \big(1 -  r^2 \big) + O(r^4),
\end{split}
\end{equation}
where we have performed the necessary calculations. The leading constant term converges quickly to $1$ for $m>7$. The coefficient for $r^2$ is defined by an alternating series. We first choose $m=3,7,15,31$ and extrapolate these results to $m \rightarrow \infty$ to get $1.0620$. We then repeat the process using $m=4,8,16,32$ to get $0.9405$. By the minimal computational effort, we verify that the coefficient does equal to unity by considering the mean value ($1.0013$). 

In fact, if the initial velocity has a more general profile,
\begin{equation}\label{hpf-gen}
w_0(r) = \sum_{i=m}^{k}\beta_m (1 - r^2)^{m+1},\;\;\;(m \geq 0,\;\;\;\mbox{finite}\;k),
\end{equation}
the time-insensitive solution can be computed in a similar manner. Consequently, we may contemplate that Poiseuille's measurements ought to be the transient of well-controlled initial parabolic profiles, which can be generated and maintained by a fixed pressure gradient during the tests.

Szyma{\'n}ski (1932) considered flow development from an impulsively started motion from rest. He derived
\begin{equation} \label{szy}
w(r,t) = -\frac{1}{4 \mu}\frac{\partial p}{\partial z}(t) \Big( \; \big (1 - r^2 \big) - 8 \sum_{k=0}^{\infty}\frac{J_0(\lambda_k r)}{\lambda_k^3 J_1(\lambda_k)} \; \exp\big(- \lambda^2_k \nu t \big) \; \Big),
\end{equation}
where it is assumed that the pressure gradient is zero for $t\leq0$, and approaches to a constant as $t>0$. The Hagen-Poseuille flow is recovered as $t \rightarrow \infty$. In reality, solution (\ref{szy}) is misleading as it implies that the approximation (\ref{hpfns-ibp}) holds for arbitrarily long time. For air and water, the duration for the unsteady effect to disappear altogether must be in the order of $t \sim O(10^6)$ seconds! On the contrary, high-Reynolds-number streamlined flows in pipes have only been observed in {\it brief periods of time} and over {\it finite space}. 
(Reynolds 1883; Pfenninger 1961; Mullin 2011), especially for fluids of small viscosity. 
 
One possible way out of the dilemma (\ref{szy}) is to adopt our idea of the limit $\nu t \rightarrow 0^+$. The first term on the right is in fact the second term in (\ref{hpf-ibp}) which is showed to be negligible in the short time limit. In practice, it is very difficult to apply a constant pressure gradient right from the beginning of the motion. The initial flow will be highly unsteady and its life span is likely too short to determine suitable modifications for subsequent repeats (so as to achieve a constant pressure gradient).  However, it is relatively easy to produce established or full-developed flows, at least over a portion of the pipe, driven by an adjustable pressure gradient. We may continuously operate our apparatus to maintain a constant flow rate for sufficiently long time. Numerous experiences show that it is crucial to control the flow conditions near the pipe inlet in order to achieve a parabolic profile over useful distance downstream. In this sense, our theory is more general than the impulsively started scenario. 
\section{Presence of swirl at pipe inlet}
Naturally, to maintain a uniform parabolic profile over the whole cross-section at the inlet is extremely difficult; the entry velocity is likely the combination of a normally parabolic distribution and a swirling component. We postulate that 
\begin{equation*}
u=v=0, \;\;\; w=w(r,\theta,t),\;\;\; \mbox{and} \;\;\; p=p(r,\theta,z,t)
\end{equation*}
over time $t \in [0,\epsilon]$. Here we have assumed that the $\theta$-dependence of the pressure is a result of accelerating the flow into the inlet, possibly from a large fluid container, so that an equivalent body force $F(r,\theta,t)$ is generated. Let $F_r$ and $F_{\theta}$ be the components in the $r$ and $\theta$ directions respectively. Thus we must have ${\partial p}/{\partial r} = F_r$, and ${\partial p}/{\partial \theta} = r F_{\theta}$.
   
The equation of motion has a simple form of
\begin{equation} \label{swirl-ns-ibp}
\begin{split}
\frac{\partial w} {\partial t} &- \nu \Big( \frac{\partial^2 w}{\partial r^2} + \frac{1}{r} \frac{\partial w}{\partial r} + \frac{1}{r^2}\frac{\partial^2 w}{\partial \theta^2} \Big) = -\frac{\partial p/\rho}{\partial z},\\
& \\
&w(r,\theta,0) = S_0(r,\theta),\;\;\; w(1,\theta,t)=0.
\end{split}
\end{equation}
The solution for the initial-boundary value problem can be written as
\begin{equation} \label{swirl-ibp}
\begin{split}
w(r,\theta,t)&= \int_0^1 \int_0^{2 \pi}  S_0(r',\theta') K(r,\theta,r',\theta',t) \rd r'\rd \theta' \\ 
&- \int_0^t \int_0^1 \int_0^{2 \pi} \Big(\frac{\partial p/\rho}{\partial z}\Big)(r',\theta',z',\tau) K(r,r',\theta,\theta',t{-}t') \rd r' \rd \theta' \rd t',
\end{split}
\end{equation}
where Green's function $K$ is given by
\begin{equation} \label{swirl-green}
\begin{split}
K(r,r',\theta,\theta',&t)= \frac{1}{\pi} \sum_{m=1}^{\infty} \exp \big({-} \lambda^2_{m}  \nu t \big)  r'\frac{J_0 \big(\lambda_{m} r \big) J_0 \big(\lambda_{m} r' \big)}{J^2_1 \big(\lambda_{m})} \\
&+\frac{2}{\pi} \sum_{n=1}^{\infty}\sum_{m=1}^{\infty} \exp \big({-} \sigma^2_{n,m}  \nu t \big)   r'\frac{J_n \big(\sigma_{n,m} r \big) J_n \big(\sigma_{n,m} r' \big)}{(J_{n+1} \big(\sigma_{n,m})\big)^2} \cos\big(n(\theta{-}\theta')\big).
\end{split}
\end{equation}
The constant $\sigma_{n,m}$ is the $m$th positive zero of $J_n(\sigma)=0$ (Watson 1944). The derivatives of the Bessel function can be evaluated from the recurrence relations 
\begin{equation*}
J'_{n}(\sigma)=J_{n-1}(\sigma)-n J_n(\sigma)/\sigma\;\;\; \mbox{and} \;\;\;
J_{n+1}(\sigma)=(2n)J_n(\sigma)/\sigma-J_{n-1}(\sigma).
\end{equation*}
Suppose that the initial data near the pipe inlet can be expressed as
\begin{equation*}
S_0(r,\theta)=\beta(1-r^2) \gamma(\theta).
\end{equation*}
In the limit of $\nu t {\rightarrow} 0$ and a constant axial pressure gradient, the leading order approximation by keeping the first sum in (\ref{swirl-green}) is proportional to the angular integral of $\gamma(\theta)$. The resulting profile remains parabolic with modified strength. For the full effects of swirl, we must include the double sum and the analysis becomes much more involved in general though it is a matter of computation.

The parabolic profile is not the only uni-directional time-insensitive flow achievable in a straight pipe. If the pressure gradient is allowed to vary as a non-linear function of $x$, a whole range of flow profiles, possibly with strong swirls, may be generated (see, for example, Pfenninger 1961). The short-time evolution of these flows can be explained by our solution (\ref{hpf-ibp}) and (\ref{swirl-ibp}) as long as we have an adequate knowledge of the initial data $w_0(r)$ and the pressure near the pipe inlet.
\section{Numerical examples}
In this section, we present some numerical results for various initial data at the entrance of a long rigid circular pipe (the radius is assumed to be unity). The profile given in solution (\ref{hpf-ibp}) is normalised by quantity $-(\partial p/\partial z)/(4 \mu)=P_0$, and we write the scaled velocity as
\begin{equation*}
w^*(r,t)=w(r,t)/P_0=w_I(r,t) + w_G(r,t),
\end{equation*}
where the first term on the right stands for the effect of scaled inlet data, $w^*_e$, and the second term is induced by the Green function at constant axial pressure gradient. 
\subsection{Initial parabolic / boundary-layer profile}
\begin{figure}[h] \centering
  {\includegraphics[keepaspectratio, height=20cm,width=12cm]{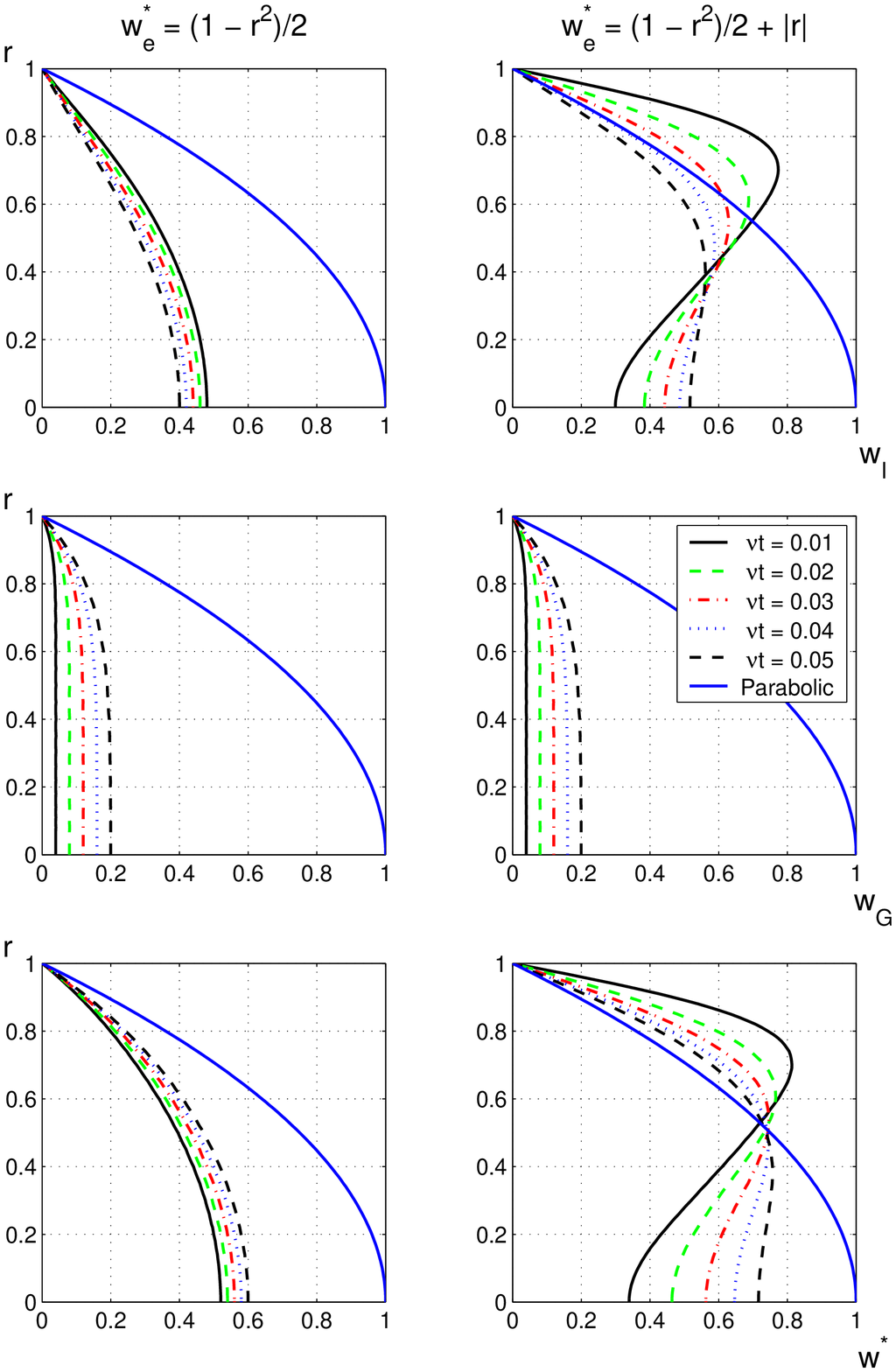}}
  \vspace{1mm}
  \caption{Flow regimes near pipe entry by uni-directional flow approximation. Entry profile is supposed to be a parabolic entry velocity or a Poiseuille-Couette profile. }\label{pprof} 
\end{figure}

\begin{figure}[h] \centering
  {\includegraphics[keepaspectratio, height=20cm,width=12cm]{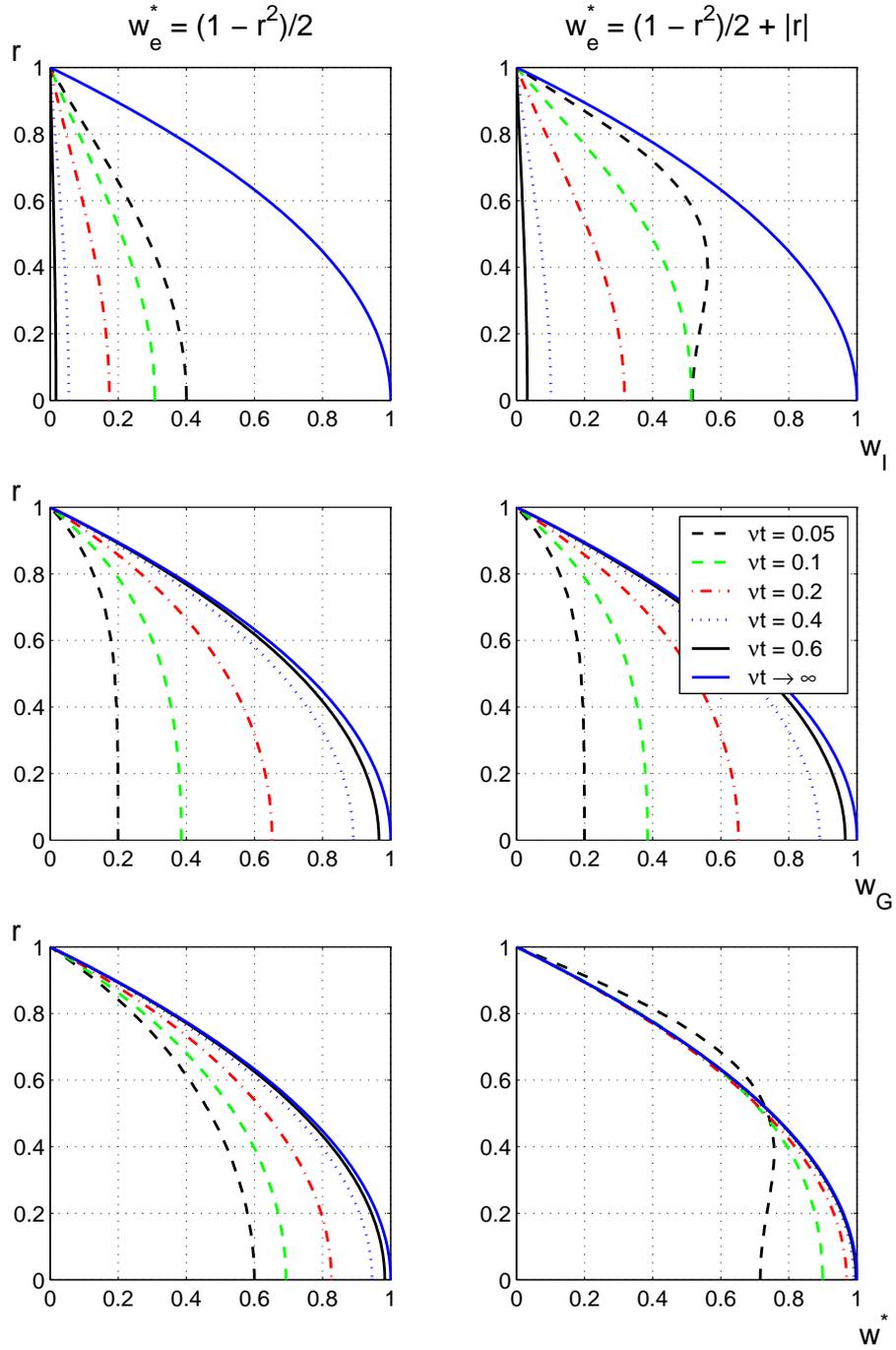}}
  \vspace{1mm}
  \caption{At large time, the velocity field driven by $\partial p/\partial z$ becomes dominant and tends to a parabolic profile in the limit of $\nu t \rightarrow \infty$. In the limiting case, it is the Green function $G(r,s,t)$ which determines the large-time decay behaviours. }\label{pprof2} 
\end{figure}

\begin{figure}[h] \centering
  {\includegraphics[keepaspectratio, height=20cm,width=12cm]{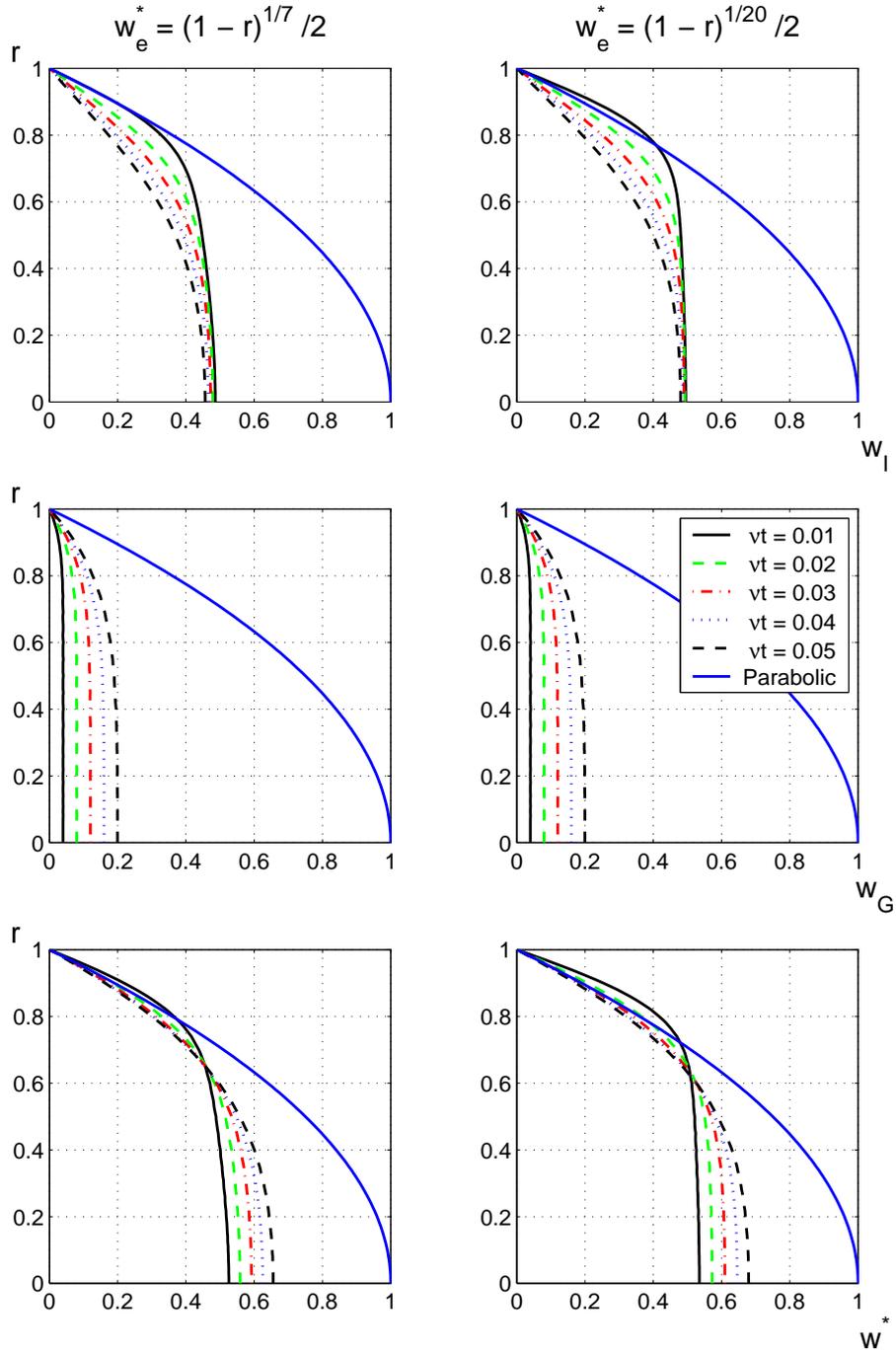}}
  \vspace{1mm}
  \caption{Flows over the pipe entry length from two boundary-layer profiles. }\label{blprof} 
\end{figure}

\begin{figure}[h] \centering
  {\includegraphics[keepaspectratio, height=20cm,width=12cm]{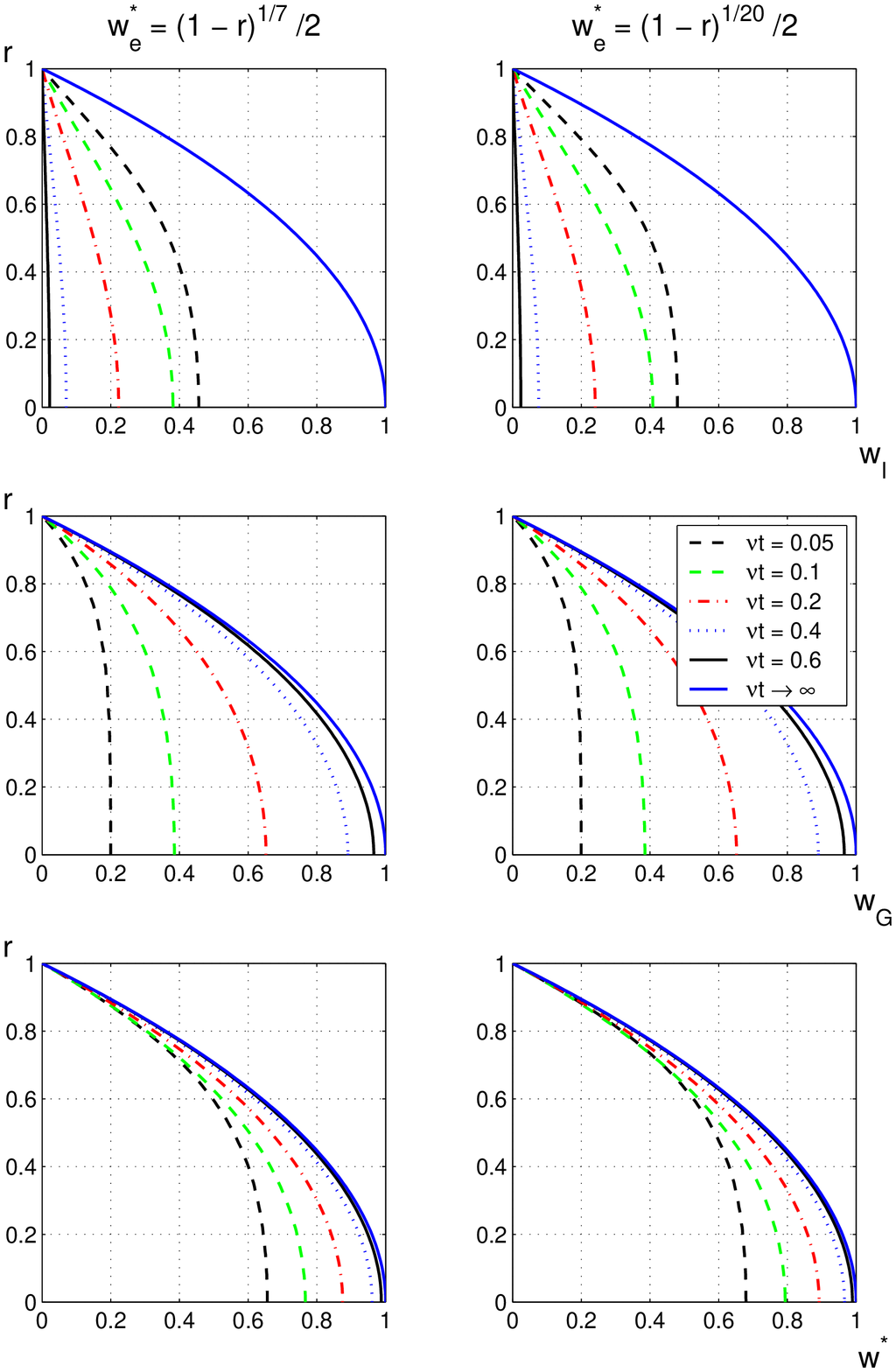}}
  \vspace{1mm}
  \caption{The initial boundary layers decay in large time.}\label{blprof2} 
\end{figure}

In figures \ref{pprof} and \ref{pprof2}, we show two examples where the initial velocity distribution is a parabolic type or a Hagen-Poiseuille-Couette type, while the pressure gradient is nominally constant to maintain a constant mass-flux through the pipe. If we assume that a motion is set up impulsively from rest, the flow development under the constant pressure gradient is in fact the solution of an impulsively started motion from a parabolic velocity obtained by Szyma{\'n}ski (1932) where the scaled profile is given by (cf. equation (\ref{szy}))
\begin{equation*}
w_G = (1-r^2) - 8 \sum_{n=1}^{\infty} \; \exp \big(- \lambda^2_n  \nu t \big) \: \frac{J_0 \big(\lambda_n r \big)}{\lambda^3_n \: J_1 \big(\lambda_n  \big)}.
\end{equation*}
This is the solution presented in the middle rows in figures \ref{pprof} and \ref{pprof2}. The computed results show that {\it initial data with parabolic distributions do not describe the flow regimes over the entry length}. Specially, the viscous layers at pipe wall are mis-represented. In addition, parabolic profiles at pipe entry are difficult to generate in practice. If the entry profiles are of boundary layer type, as shown in figures \ref{blprof} and \ref{blprof2}, the developed velocity downstream of the entry resembles the well-known experimental measurements (see, for example, \S2.5 of Tritton 1988). A parabolic velocity profile can only be realised in the limit of $\nu t \rightarrow \infty$ or at locations far downstream of the entry as long as the axial pressure variation is kept constant.
\subsection{Entry profile with swirl}
\begin{figure}[h] \centering
  {\includegraphics[keepaspectratio, height=18cm,width=14cm]{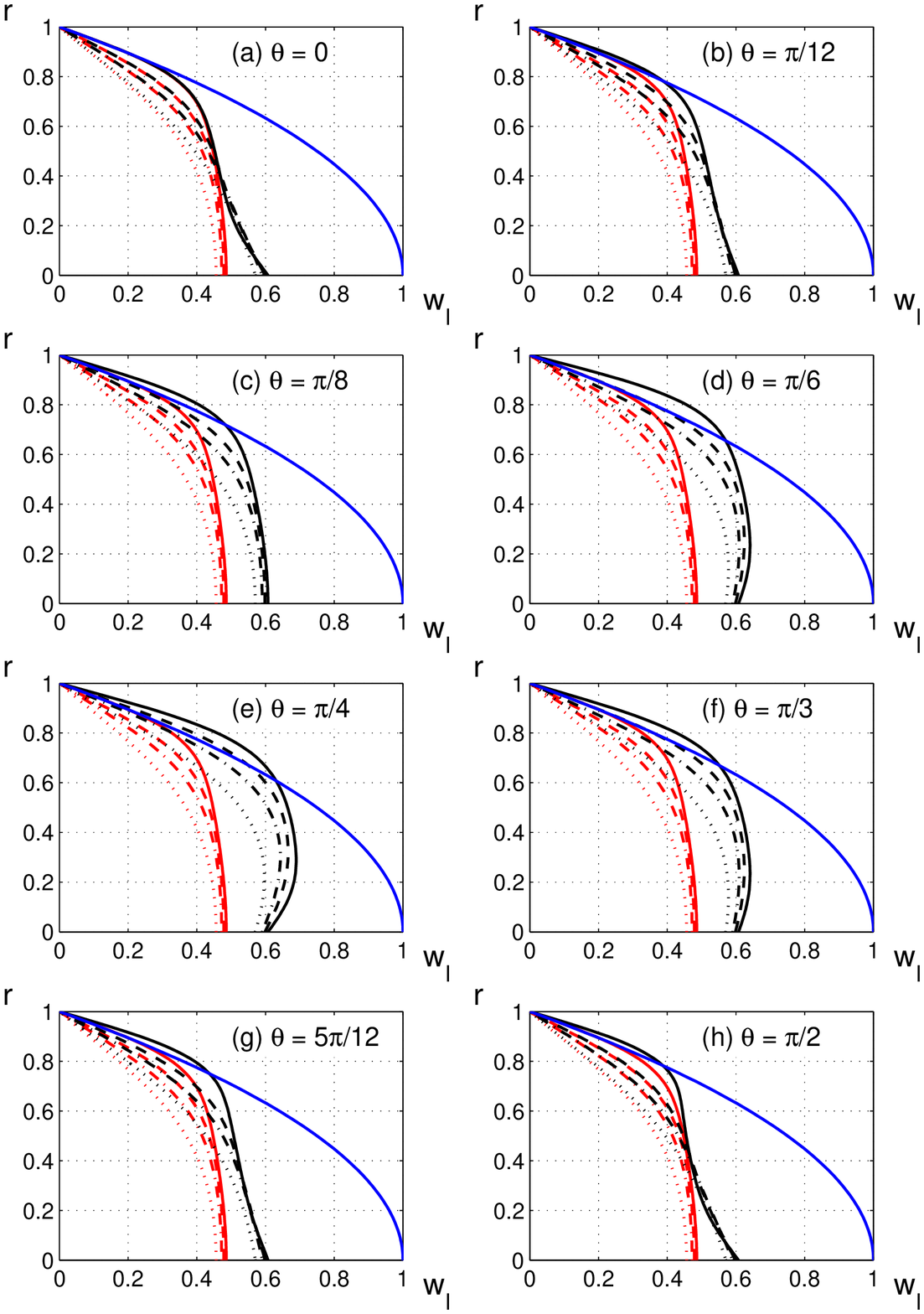}}
  \vspace{1mm}
  \caption{Velocity profiles at $8$ angular locations are shown in plots (a) to (h) for $\nu t=0.01$, $0.02$, $0.03$, $0.05$ (black lines) compared with parabolic profile and boundary layer without swirl (red lines).  }\label{swprof} 
\end{figure}

We show one example of inlet flow with swirl with the natural period in the azimuthal direction:
\begin{equation*}
w^*_e(r,\theta) = \frac{1}{2}(1-r)^{1/7} \Big( 1 + \frac{1}{2} \sin^2 (2 \pi \theta) \Big)
\end{equation*}
As assumed in the present theory, the swirl is generated by an body force near the pipe entrance and the subsequent flow has a constant mass-flux or a constant axial pressure gradient. Because of the $\theta$ rotational symmetry in Green's function (\ref{swirl-green}), the second contribution in (\ref{swirl-ibp}) is identical to the non-swirl cases just examined. Hence only the computational results of $w_I$ are presented in figure \ref{swprof} where, clearly, the overall flow field has certain "symmetries" and is substantially more complicated to analyse. We shall carry out detailed investigation in separate studies.
\section{Conclusion}
We have shown that, by including $\partial u/\partial t$ term in the equations of motion, the uni-directional flow in a channel (\ref{ppf-ibp}) or in a pipe (\ref{hpf-ibp}) is {\it space-time-dependent} and exists over a short time when the equations are solved with {\it non-zero} initial data. The classical plane Poiseuille flow (\ref{ppf}) or Hagen-Poiseuille profile (\ref{hpf}) are best interpreted as an evolutional solution of an appropriate initial velocity in the limit of $\nu t \rightarrow 0^+$. This point has been heuristically argued by Lam (2013) on the basis of exact vorticity solutions in ${\mathbb R}^3$. The subsequent development over longer periods must be governed by the full equations of motion where the non-linearity becomes dominant. It has long been known that, by neglecting the non-linear term $(u.\nabla)u$, the approximate fluid motions are only valid for low Reynolds numbers, see, for example, \S 110 of Prandtl \& Tietjens (1934). 

It is not difficult to understand why there do not exist absolute steady flows, independent of initial data, at arbitrary Reynolds numbers. Our flow regimes are clearly outside the physics of relativity, the concept of space-time entity enables us to recognise the conceptual inconsistency in the formulas (\ref{hpf}) or (\ref{ppf}). 

Although the differences in the spatio-temporal solution (\ref{hpf-ibp}) and the absolute steady flow (\ref{hpf}) appear to be insignificant with reference to experimental verifications, particularly for fluids of small viscosity, the hydrodynamic principles underlying the two fluid motions have important ramifications. For instance, the linear stability analysis for the non-dimensional parabolic profile shows that the basic flow is stable to all infinitesimal disturbances (see, for example, Lam 2014). By this counter-example, one casts serious doubts on the relevance of the linear stability theory in explaining the process of laminar-turbulent transition. Could it be something fundamental which may have gone wrong? If the basic flow evolves in space {\it and} time, what is the meaning of its stability or instability when the Reynolds number itself is varying?

\vspace{10mm}
\begin{acknowledgements}

\noindent 
04 May 2015

\noindent 
\texttt{f.lam11@yahoo.com}
\end{acknowledgements}
%
%
\addcontentsline{toc}{section}{\noindent{References}}

\label{lastpage}
\end{document}